\documentclass[12pt,preprint]{aastex}

\def\lapp{\ifmmode\stackrel{<}{_{\sim}}\else$\stackrel{<}{_{\sim}}$\fi}
\def\gapp{\ifmmode\stackrel{>}{_{\sim}}\else$\stackrel{>}{_{\sim}}$\fi}

\begin{document}

\title{Timing and Flux Evolution of the Galactic Center Magnetar SGR~J1745$-$2900}

\author{
Victoria M. Kaspi\altaffilmark{1,2}, 
Robert F. Archibald\altaffilmark{1},
Varun Bhalerao\altaffilmark{3},
Fran{\c c}ois Dufour\altaffilmark{1},
Eric~V.~Gotthelf\altaffilmark{4},
Hongjun An\altaffilmark{1},
Matteo Bachetti\altaffilmark{5,6}, 
Andrei M. Beloborodov\altaffilmark{4}, 
Steven E. Boggs\altaffilmark{7},
Finn E. Christensen\altaffilmark{8}, 
William~W.~Craig\altaffilmark{7,9},
Brian W. Grefenstette\altaffilmark{10}, 
Charles J. Hailey\altaffilmark{4},
Fiona A. Harrison\altaffilmark{10}, 
Jamie A. Kennea\altaffilmark{11},
Chryssa Kouveliotou\altaffilmark{12}, 
Kristin K. Madsen\altaffilmark{10}, 
Kaya Mori\altaffilmark{4},
Craig B. Markwardt\altaffilmark{13},
Daniel Stern\altaffilmark{14}, 
Julia K. Vogel\altaffilmark{9}, 
and William W. Zhang\altaffilmark{13}\\}
\affil{
{\small $^1$Department of Physics, McGill University, Montreal, Quebec, H3A 2T8, Canada}\\
{\small $^2$vkaspi@physics.mcgill.ca}\\
{\small $^3$Inter-University Center for Astronomy and Astrophysics, Post Bag 4, Ganeshkhind, Pune 411007, India}\\
{\small $^4$Columbia Astrophysics Laboratory, Columbia University, New York NY 10027, USA}\\
{\small $^5$Universit{\'e} de Toulouse, UPS-OMP, IRAP, Toulouse, France}\\
{\small $^6$CNRS, Institut de Recherche en Astrophysique et Plan{\'e}tologie, 9 Av. colonel Roche, BP 44346, F-31028 Toulouse Cedex 4, France}\\
{\small $^7$Space Sciences Laboratory, University of California, Berkeley, CA 94720, USA}\\
{\small $^8$DTU Space, National Space Institute, Technical University of Denmark, Elektrovej 327, DK-2800 Lyngby, Denmark}\\
{\small $^9$Lawrence Livermore National Laboratory, Livermore, CA 94550, USA}\\
{\small $^{10}$Cahill Center for Astronomy and Astrophysics, California Institute of Technology, Pasadena, CA 91125, USA}\\
{\small $^{11}$Department of Astronomy \& Astrophysics, The Pennsylvania State University, 525 Davey Lab, University Park, PA 16802, USA}\\
{\small $^{12}$Astrophysics Office, ZP 12, NASA Marshall Space Flight Center, Huntsville, AL 35812, USA}\\
{\small $^{13}$Goddard Space Flight Center, Greenbelt, MD 20771, USA}\\
{\small $^{14}$Jet Propulsion Laboratory, California Institute of Technology, Pasadena, CA 91109, USA}\\
}

\begin{abstract}

We present the X-ray timing and spectral evolution of the Galactic Center magnetar
SGR J1745$-$2900 for the first $\sim$4 months post-discovery using data obtained with the 
{\it Nuclear Spectroscopic Telescope Array (NuSTAR)} and {\it Swift} observatories.
Our timing analysis reveals
a large increase in the magnetar spin-down rate by a factor of $2.60 \pm 0.07$ over
our data span.  We further show that the change in spin evolution was likely coincident with a bright
X-ray burst observed in 2013 June by {\it Swift}, and if so, there was no accompanying discontinuity
in the frequency.   We find that the source
3--10 keV flux has declined monotonically by a factor of $\sim$2 over an 80-day period post-outburst
accompanied by a $\sim$20\% decrease in the source's blackbody temperature, although there is evidence
for both flux and $kT$ having levelled off.  We argue that the torque variations are likely to be
magnetospheric in nature and will dominate over any dynamical signatures of orbital motion
around Sgr A*.

\end{abstract}

\keywords{stars: neutron --- stars: magnetic field --- pulsars: general --- X-rays: stars}

\section{Introduction}

The recently identified Galactic Center magnetar SGR J1745$-$2900 has a brief but
interesting observational history.  
It was discovered serendipitously during an ongoing monitoring program of the Galactic Center region with the Swift X-ray Telescope (XRT). On 2013 April 24 increased X-ray emission was detected from the SGR A* region \citep{drm+13}, followed the next day by a bright X-ray burst reported by Swift's Burst Alert Telescope 
(BAT) \citep{kkb+13,kbk+13}. 
{\it Swift} XRT observations
that same day refined the position of the burster to within 2.8$''$ of Sgr A* \citep{kkb+13}.
Target-of-Opportunity observations by 
the {\it Nuclear Spectroscopic Telescope Array (NuSTAR)} revealed
3.76-s pulsations from the new transient, and measured a spin-down rate
that implies the presence of a neutron star having surface equatorial dipolar
magnetic field strength\footnote{Estimated assuming simple magnetic
braking in a vacuum via $B=3.2 \times 10^{19} \sqrt{P \dot{P}}$~G.}
$1.6 \times 10^{14}$~G \citep{mgz+13}.  This identified the source as a
newly outbursting magnetar in the Galactic Center (GC) region.  
\citet{mgz+13} also showed that the source spectrum was well described
by a blackbody of $kT=1$~keV plus a power law of index of 1.5.
A {\it
Chandra} observation later confirmed the GC association and localized the source to
an offset from Sgr A* of only 2.4$''$ \citep{rep+13}.  \citet{ekk+13} and
\citet{sj13} reported on the detection of the radio pulsar counterpart, and
\citet{efk+13} showed that the observed value of the rotation measure of the radio
pulsar constrains the strength of the magnetic field near Sgr A*,
which provides a unique test of radiative accretion theory for supermassive black holes. 

\citet{mgz+13}
asserted that the spin-down rate of the magnetar is sufficiently large
that bias due to dynamical effects in the
GC region will be negligible, unless the measured spin-down rate were temporarily greatly
enhanced \citep[e.g. due to glitch recovery; see][]{dkg08}. 
\citet{rep+13} argued dynamical effects may in principle be measurable at the $\sim$10\% level
with long-term monitoring.  However, the latter would have to be in spite of
the likely continued fading of the source back to quiescence, as well as
the often highly noisy nature of magnetar spin evolution post-outburst
\citep[e.g.][]{wkg+02,gk04,dkg09,crj+08,dksg12}.

Also of interest, independent of the Galactic Center location, is the magnetar outburst
itself.  Specifically, the flux and spectral evolution of magnetars post-outburst
can potentially constrain the physics of neutron star magnetospheres
and/or crustal and interior composition.
In the former case, magnetar outbursts are hypothesized to be due to twists in localized
magnetospheric regions of enhanced current known as ``j-bundles'' \citep{bel09}.  Untwisting
of j-bundles involves the return of current to a hot spot on the stellar surface, with gradually
decreasing luminosity and temperature, predictions that can be tested by measuring flux and
spectral evolution post-outburst.
In the latter case, models of crustal cooling following a sudden heat injection can be
fit to magnetar cooling curves, and can constrain e.g. the depth of the energy injection
as well as the nature of the stellar temperature profile \citep[e.g.][]{kew+03,snl+12,pr12,akac13}.
In either case, a significant hardness/flux correlation is expected and indeed thus far
is generally observed \citep[e.g.][]{re11,sk11}.

Here we report on continued {\it NuSTAR} and {\it Swift} XRT monitoring of SGR J1745$-$2900
post-outburst, specifically its timing and flux evolution.  We show that the source's
spin-down rate has recently undergone a large increase in magnitude, by over a factor of two.
We suggest that the change in rate occurred coincidentally with a second X-ray burst seen by {\it Swift} BAT
on MJD 56450 (7 June 2013) \citep{kbc+13}.
If the burst association is correct, this change in spin-down rate occurred with no coincidental
period glitch and without a large radiative change beyond the short $<$0.32-s burst and
possibly slightly elevated flux on that day as reported by \citet{kbc+13}.  
We also report on the source's flux and spectral evolution $>$100 days post-outburst.
%

\section{Observations, Analysis and Results}

The {\it NuSTAR} mission consists of two co-aligned focusing X-ray telescopes operating
in the range 3--79 keV \citep{hcc+13}.  X-rays are focused onto CdZnTe chips (4 chips
for each of two modules, A and B), yielding a point-spread function of FWHM $\sim$18$''$. 
The {\it NuSTAR} detectors have 2-ms time resolution, more than adequate for studying 
the 3.76-s pulsar SGR J1745$-$2900.
{\it NuSTAR} observed SGR J1745$-$2900 a total of 13 times between
MJDs 56408 (2013 April 26) and 56517 (2013 August 24) with integration
times listed in Table~1.

\subsection{Timing}
\label{sec:timing_results}

\begin{center}
\begin{figure}[t]
\epsscale{0.75}
\plotone{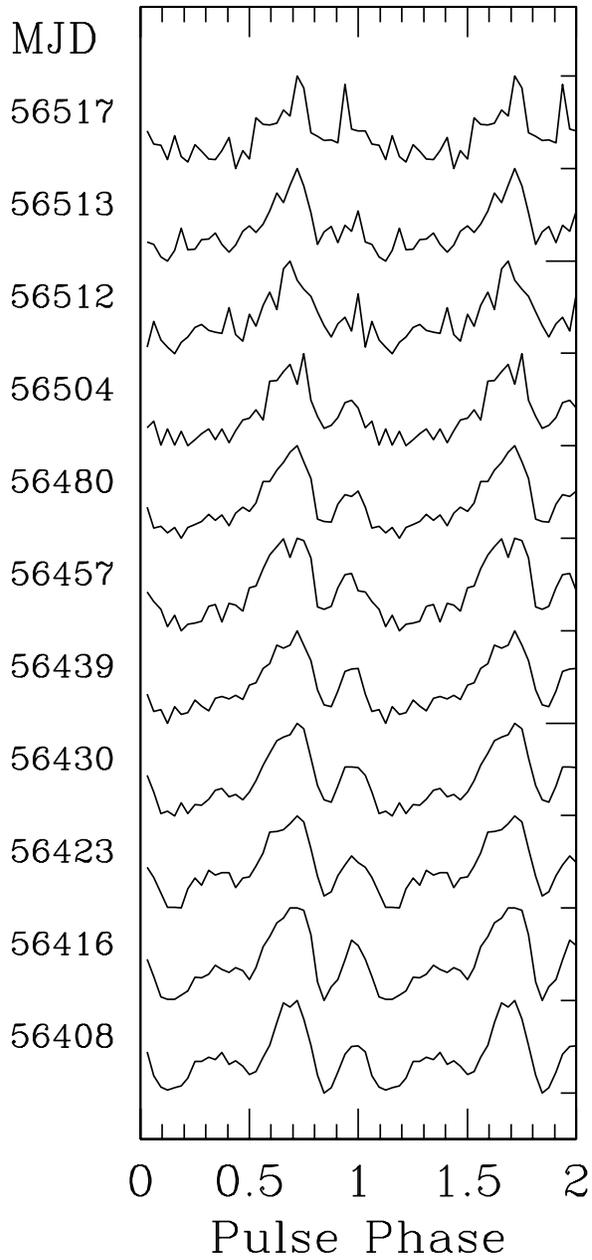}
\figcaption{
{\it NuSTAR} pulse profiles in the 3--10~keV band at
the observing MJDs (see Table 1), aligned using the
ephemerides presented in Table~\ref{ta:timing}.  Two cycles are
shown for clarity.  Note the
gradual disappearance of the first of the three peaks
seen in the MJD 56408 observation.  Uncertainties on
phase bins are omitted for clarity but are well represented
by the off-pulse scatter.
\label{fig:profs}
}
\end{figure}
\end{center}

For timing purposes, for each {\it NuSTAR} observation, we extracted photons in a 1$'$ radius around
the nominal source position using the {\it NuSTAR Data Analysis Software (NuSTARDAS)} version 1.2.0,
along with {\tt HEASOFT} version 6.13. The data were reduced to
the Solar System barycenter assuming the {\it Chandra} position reported by \citet{rep+13} and
the DE200 planetary ephemeris.
We also filtered the
data to extract only photons with energies 3--10~keV, as this generally provided
an optimal signal-to-noise strength on the pulse.  Average pulse times-of-arrival (TOAs) were extracted
from the event lists by first folding the X-ray time series at the nominal pulse period and then
aligning the resulting profile with a high signal-to-noise ratio template in the Fourier domain taking
into account 6 Fourier harmonics, although our results are not strongly dependent on
this choice.  We have observed the
pulse profile to be largely stable long-term,
apart from the slow disappearance of the first peak seen in Figure~1 of \citet{mgz+13}; 
see our Figure~\ref{fig:profs}.
This gives us confidence in the reliability of the TOA extraction method as the primary
and third peaks have remained unchanged.  The resulting
TOAs were then fed into the {\tt tempo} software package\footnote{http://www.atnf.csiro.au/research/pulsar/tempo/} for further analysis.
Note that we have verified that {\it NuSTAR}'s clock is sufficiently stable (i.e. reliable in absolute timing to well 
under 10 ms on comparable time scales to those considered here) that it
contributes negligibly to the uncertainties in the TOAs.

We further supplemented the {\it NuSTAR} TOAs with timing data extracted from {\it Swift} XRT \citep{bhn+05}
Windowed Timing (WT) observations of the source.  Photon-counting mode
data could not be included as they had insufficient timing resolution.
To produce the {\it Swift} TOAs, Level 1 data products were
obtained from the HEASARC {\it Swift} archive and reduced using the
$xrtpipeline$ standard reduction in HEASOFT version $6.13$.   We extracted
photons in a 
47$''$ radius around the nominal source position, and
reduced the resulting event times to the Solar System barycentre. The
{\it Swift} data were also filtered to include only photons from 3--10 keV.
The resulting events
were then subjected to the same TOA extraction analysis as for the {\it NuSTAR} data.
The details of the {\it Swift} observations are presented in Table~1.
Note that one WT-mode observation (ObsID 00032811004 on MJD 56429) was omitted from
the analysis due to poor statistics.

The ephemeris reported by \citet{mgz+13} provided an excellent fit to the initial
{\it NuSTAR} data, and we further refine it here with subsequent observations.
Our best-fit parameters for this first ephemeris are presented in Table~\ref{ta:timing}.
However TOAs added from observations made on MJDs 56439 and 56457 deviated significantly by $\sim$0.1
from the prediction of an ephemeris fit using all earlier data.  This deviation alone
is not large enough to rule out extrapolation of this ephemeris to those epochs, as
their phase deviation could be mostly fit out using a large second frequency derivative.
By MJD 56480, however, this first ephemeris clearly described the phase data poorly, precluding
proper phase counting, even with a second derivative.

\begin{figure}[t]
\epsscale{0.85}
\plotone{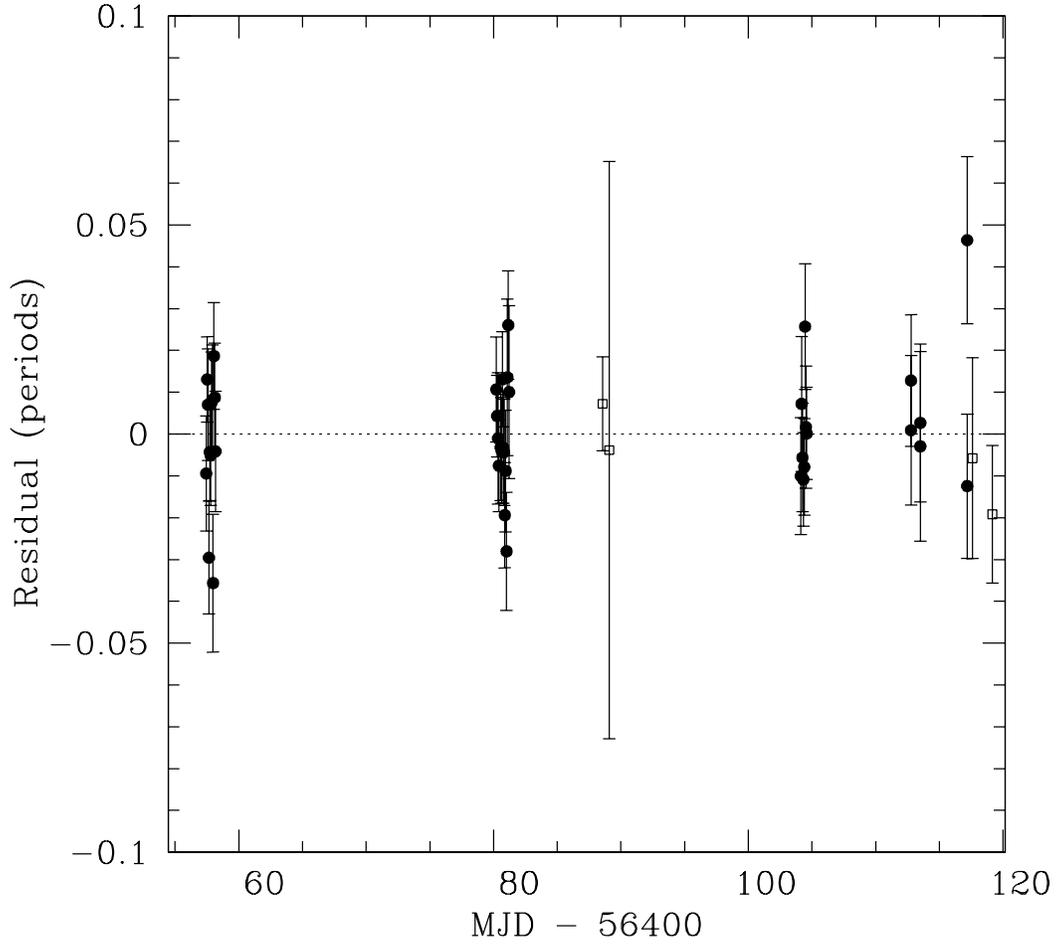}
\figcaption{
Residuals for the second phase-coherent timing solution
shown in Table~\ref{ta:timing} for the appropriate MJD range.
Filled circles are {\it NuSTAR} TOAs and empty squares are {\it Swift} TOAs.
\label{fig:resid}
}
\end{figure}

\begin{figure}[t]
\epsscale{0.85}
\plotone{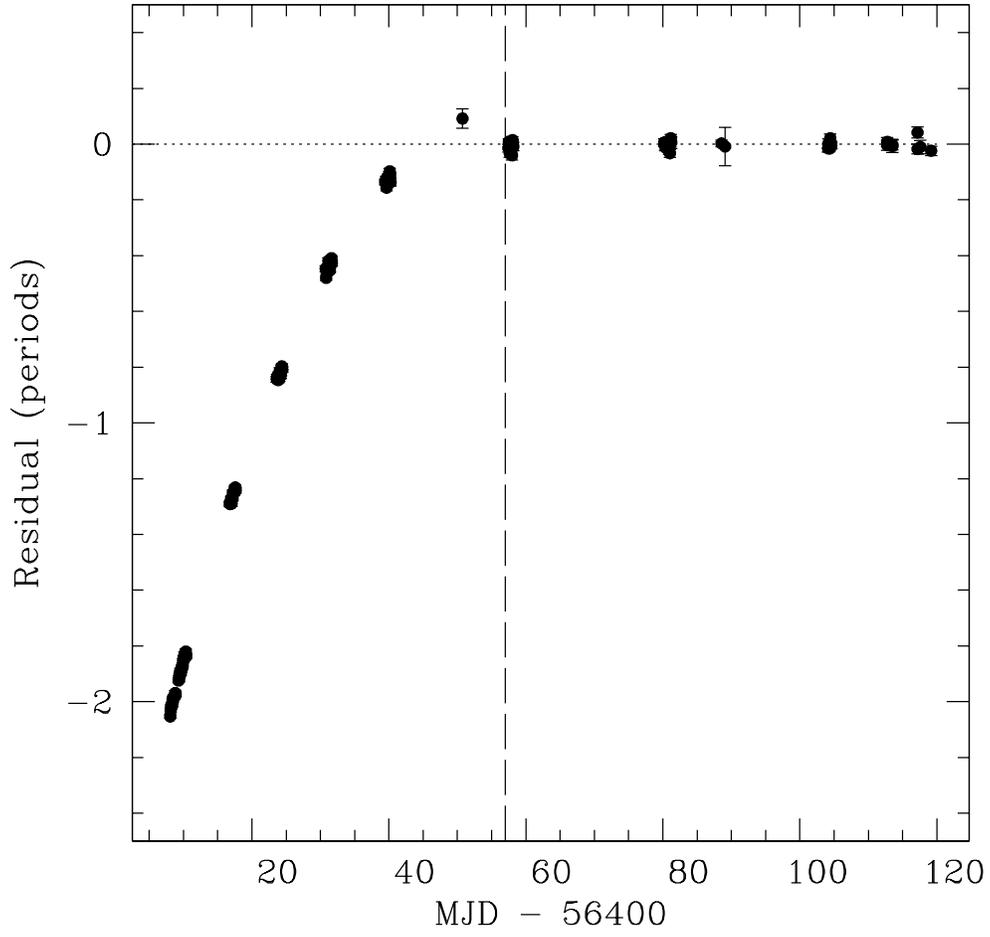}
\figcaption{
Attempt to extrapolate the second timing solution backward,
showing growing phase deviations that demonstrate the change
in ephemeris. The start of the 2nd ephemeris is indicated by
the vertical dashed line.
\label{fig:residual}
}
\end{figure}

We therefore initiated a series of closely spaced observations in order to reacquire
phase lock.  This resulted in a second phase-coherent ephemeris with significantly
different spin-down rate, as shown in Table~\ref{ta:timing}; the featureless residuals
from this new ephemeris are shown in Figure~\ref{fig:resid}.  The backward extrapolation
of this second ephemeris to MJDs 56457 and 56439 also showed a significant phase
deviation, however again a (new) second frequency derivative could reasonably be fit to
remove the deviation.  Backward extrapolation of the new ephemeris beyond those
epochs resulted in significant phase wraps, as shown in Figure~\ref{fig:residual}, even with
this new second derivative.

\begin{figure}
\epsscale{1.0}
\plotone{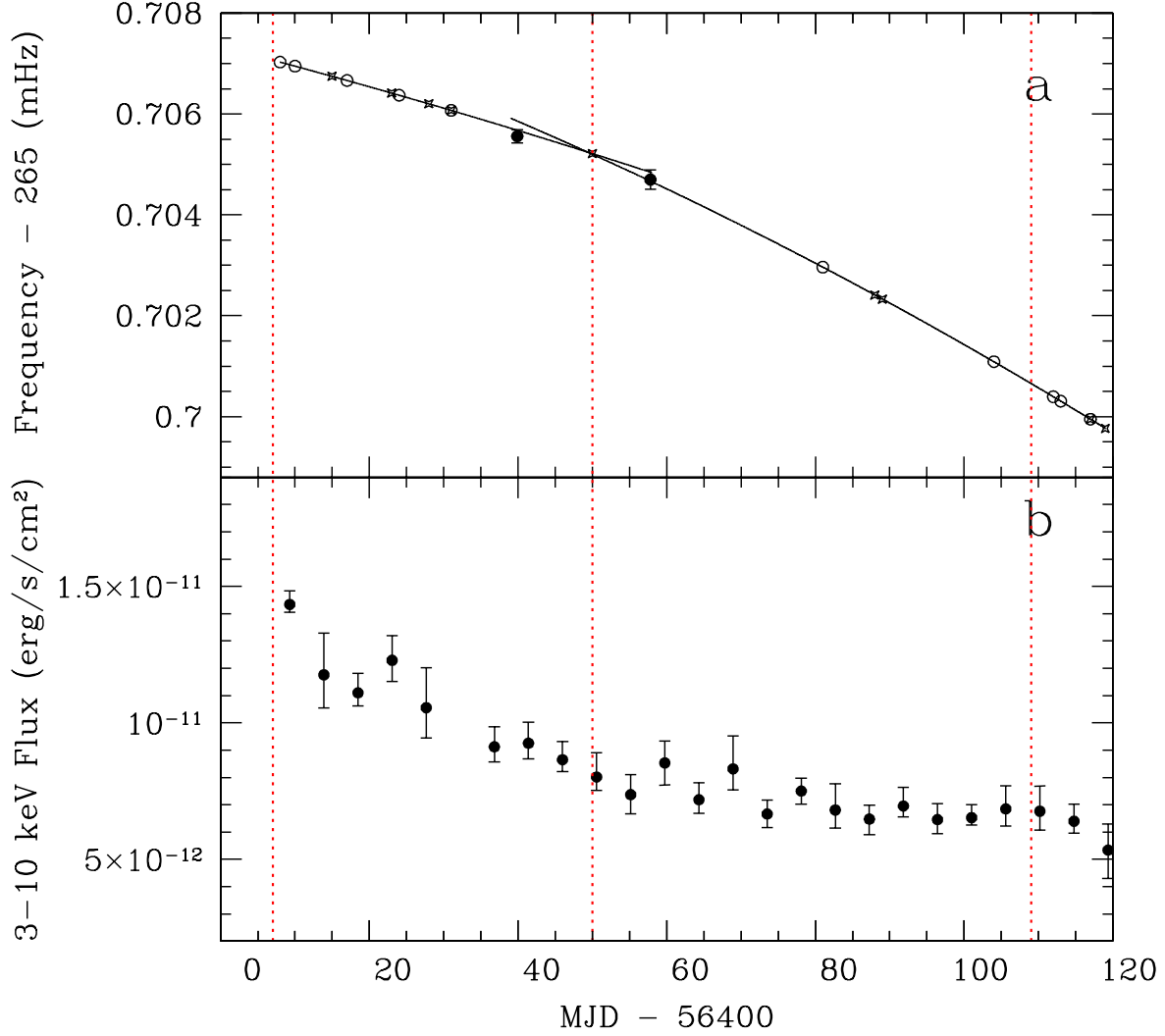}
\vspace{-0.5in}
\figcaption{\small
(a) 
Spin frequency versus time.  The two solid lines represent the two different
phase-coherent solutions discussed in \S\ref{sec:timing_results} and presented
in Table~\ref{ta:timing}.  Open circles and stars represent epochs of {\it NuSTAR} 
and WT-mode {\it Swift} XRT observations, respectively
(see Table~1) with the frequencies calculated from the phase-coherent
ephemerides.  Note the overlap region between {\it NuSTAR} observations on
MJDs 56439 and 56457 where both solutions can reasonably fit the phase data;
locally calculated frequencies for those data sets are shown in solid circles
along with error bars.  The dotted
vertical red lines indicate epochs of {\it Swift} BAT reported bursts
on MJDs 56407 (April 25), 56450 (June 7) and 56509 (August 5).
(b)  Absorbed 3--10 keV flux versus time in 5-day averages from PC-mode {\it Swift} XRT observations. 
The gap in coverage near MJD 56430 was when nearby Transient 1 contaminated
the magnetar fluxes \citep{dkg+13,dwr+13}.
\label{fig:freq}
}
\end{figure}

The two coherent ephemerides are plotted in solid lines in Figure~\ref{fig:freq}.
There the overlap region between MJDs 56439 and 56457 is shown and the difference
between the two ephemerides is clear.  To determine which ephemeris better fits the
data in the overlap region, we fit local
frequencies to the TOAs at those epochs, as the {\it NuSTAR} integration times
were sufficiently long to allow this measurement.  The resulting frequencies are plotted in solid circles,
with {\tt tempo}-reported 1$\sigma$ uncertainties shown.  Clearly the frequency
from the MJD 56439 observation is inconsistent with the second ephemeris, indicating
those data are best described by the first ephemeris. 

In Figure~\ref{fig:freq} we plot in red the epochs of the three X-ray bursts reported
by {\it Swift} BAT \citep{kkb+13,kbc+13,kbc+13b}.  
Interestingly, the second burst coincided within uncertainties
to the epoch at which our two ephemerides converge; specifically,
at the observed burst epoch (11:17:26 UT on June 7, 2013, or
MJD 56450.47044), the extrapolated frequencies of the two ephemerides agree
at the 1.9$\sigma$ level.  
This suggests that the burst likely
coincided with the change in spin-down rate, if indeed the change was abrupt.  It further
suggests that the burst event
occurred with no frequency discontinuity, i.e. with no spin-up or spin-down glitch.
We set a $3\sigma$ upper limit on the amplitude of such a glitch, assuming it coincided with
the BAT burst, of $\Delta\nu/\nu < 1.1 \times 10^{-6}$.  
This upper limit is in the mid-range of observed fractional frequency changes in magnetar
bursts \citep[e.g.]{kgw+03,dkg08,dkg09}.
If, in fact, the burst was not coincident with the ephemeris
change, then the latter accompanied a spin-up glitch if it preceded the burst, and a
spin-down glitch if it followed.  On the other hand, the change in ephemeris may
have been gradual and with no frequency discontinuity; in that case, however, that the burst epoch coincided with the
convergence of the two independently determined ephemerides would have to be merely luck.

\subsection{Flux and Spectroscopy}
\label{sec:spectral}

\subsubsection{\it Swift Observations}
\label{sec:swift}

First, to consider the overall soft-band flux evolution of the source, we analyzed Photon Counting
mode data from {\it Swift} XRT. 
Specifically, we include 109 PC-mode observations obtained between MJDs 56407 and 56550.
For this work we did not use WT-mode data as they suffered from very high background and
were not informative, but we verified they were broadly consistent with the PC mode results.
To extract the {\it Swift}  fluxes, we
obtained Level 1 data products from the {\tt HEASARC} {\it Swift} archive and reduced them
using the xrtpipeline standard reduction in {\tt HEASOFT}, using grade 0 data,
and including an exposure map. The selected source region is a circle with
20$''$ radius centered at the {\it Chandra} position for SGR J1745$-$2900. This
radius was selected as it approximates the  {\it Swift}  XRT half-power
diameter at 4 keV (Moretti et al. 2005). A source-free background
region of the same size was selected in a nearby region. Observations
were typically 1-ks long, and occurred nearly daily. 
Spectra from the observations were summed in 5-day intervals and the
results grouped with a minimum of 3 counts per spectral bin.
Spectra were fit using the `lstat' statistic, and absorption modelled
using \citet{wilms} abundances and \citet{verner} cross sections.
Fluxes measured between MJD 56430-35 were contaminated by a transient source
(Transient 1; see \S\ref{sec:nustar}) so were omitted from the analysis.
The spectra were fit linking $N_H$ as well as $kT$, as this was statistically
preferred, i.e. a variable $kT$ did not improve the quality of the fit significantly,
given the available statistics.  
In this way, we found best-fit values of $N_H=15.3^{+0.7}_{-0.6} \times 10^{22}$~cm$^{-2}$ and $kT=0.94\pm0.02$~keV.
These value are consistent with the results of Mori et al. (2013) as well as with those from {\it Chandra}
\citep{rep+13}.
The flux evolution that results from these fits is shown in Figure~\ref{fig:freq}. 


\subsubsection{\it NuSTAR Observations}
\label{sec:nustar}

The analysis of the {\it NuSTAR} spectral data for SGR J1745$-$2900 required particular
care because 
two nearby transient sources impacted the data at different times.
These transients are
CXOGC J174540.0$-$290005, 24$''$ from the magnetar (henceforth Transient 1; see ATELs 5095, 5074)
and AX J1745.6$-$2901, 88$''$ away (henceforth Transient 2; see ATELs 5226, 1513).
Both transients are Low Mass X-ray Binaries (LMXBs) and contaminated
the {\it NuSTAR} magnetar spectral data significantly; the contamination
was so severe we ignored
the spectral data in the epochs when Transient 1 was bright (MJDs 56430, 56439),
and processed with great caution when Transient 2 was bright (MJDs 56504, 56512, 56513, 56517).
Stray light from an unrelated source also contaminated the background in Module
B for one observation; see Table 1.
Data were processed
using \texttt{nupipeline} and \texttt{nuproducts} from the public release
with {\tt HEASOFT}~6.14. The extraction region was selected as a circle of
30\arcsec\ radius centered on the source position. 
Response matrices appropriate 
for each data set were also generated using the standard software.
In our spectral fitting, for interstellar X-ray absorption modelling, we assumed
\citet{wilms} abundances and \citet{verner} cross sections.
We considered only the energy range 3--30 keV, as the source was not
detected at higher energies in any observation.

The Galactic Center region is crowded and the background is spatially
variable. Moreover, the unrelated transients affected the data significantly.
As a result, extracting a separate background region for
the magnetar was in general not feasible.  In order to evaluate the robustness
of our results,
we used two different, independent methods to analyse these data.
The first method subtracted background spectra obtained from
pre-outburst data, while the second modelled the background
spectra along with the magnetar's.  
For all the observations under consideration,
in the source extraction region around the magnetar and in the energy band 
we analyzed (3--30 keV), Galactic center diffuse emission 
dominated while the internal and stray-light backgrounds were negligible. 
Both of our methods described below are appropriate when the background is dominant.
As we show, the two methods largely agree, implying our reported results are robust.

In Method 1, we selected one of the pre-outburst
images of the field (ObsID 30001002003 taken on MJD 56413) as our
background exposure. The integration time for this observation was much longer than in any of the
magnetar exposures, so the uncertainties in fit parameters are dominated by source
statistics.
We have verified that the off-axis angle of the source
(and pre-outburst background) was similar for all epochs so 
that variations in the ARF between source and background regions are small
(under $\sim$5\%).

\begin{figure}
\epsscale{0.65}
\plotone{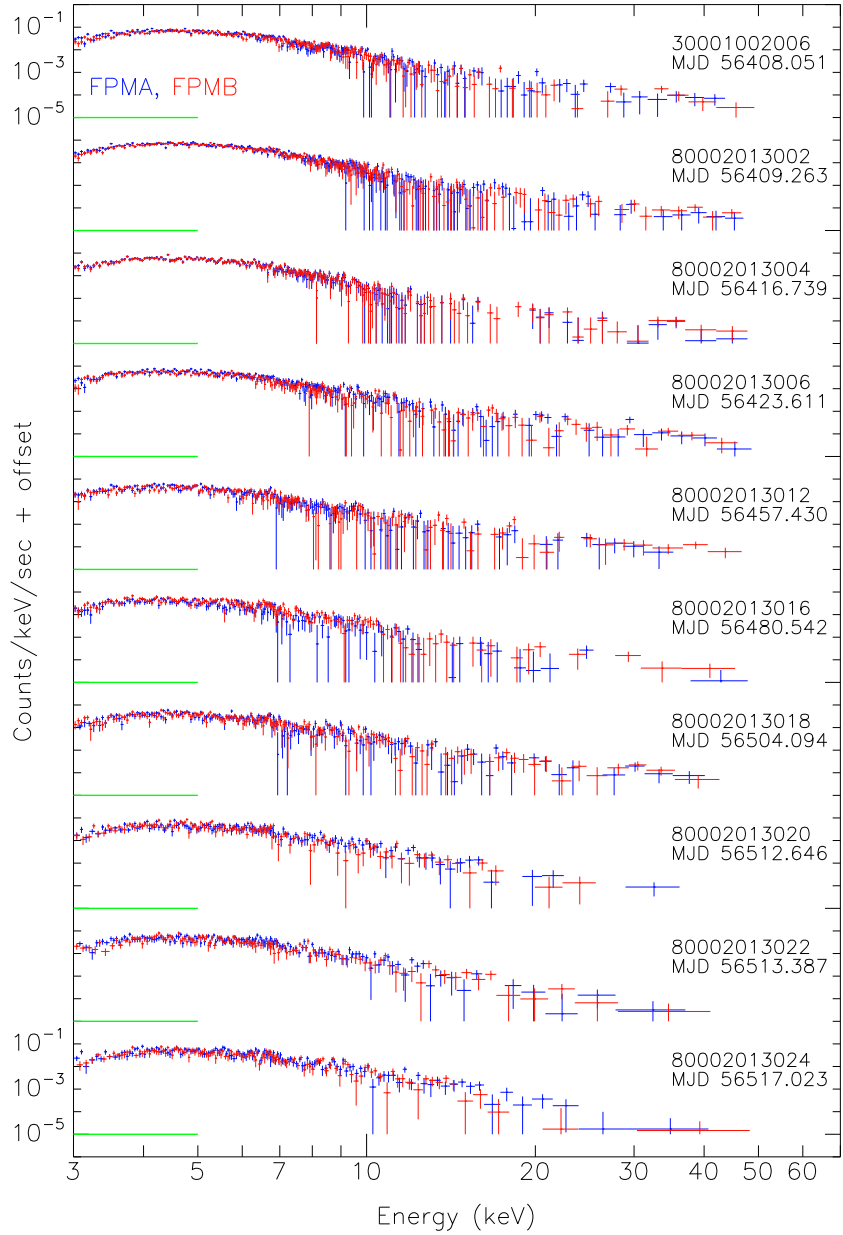}
\vspace{0.25in}
\figcaption{\small
Spectra for all {\it NuSTAR} observations from earliest (top) to latest (bottom).
FPMA is plotted in blue and FPMB is plotted in red.  Spectra have been grouped
to have a minimum of 20 counts per spectral bin.
\label{fig:all_spectra}
}
\end{figure}

\begin{figure}
\epsscale{0.95}
\plotone{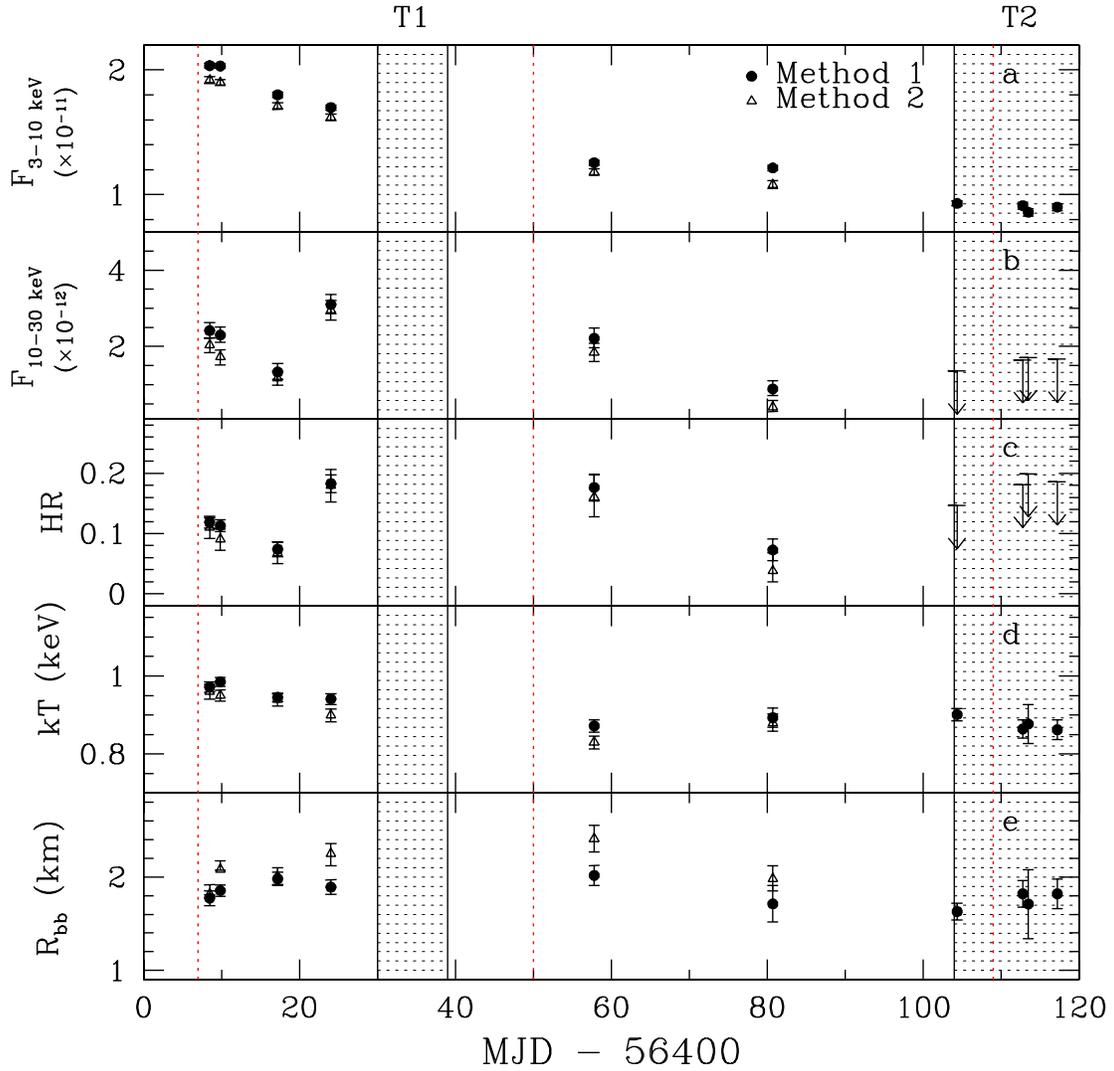}
\figcaption{\small
(a) Absorbed flux versus time in the 3--10~keV band in erg~s$^{-1}$~cm$^{-2}$, for Method 1 (solid circles)
and Method 2 (open triangles).
(b) Absorbed flux versus time in the 10--30~keV band in erg~s$^{-1}$~cm$^{-2}$, for Methods 1 
and 2.  Note that when Transient 2 is on, we present only upper limits for Method 1 
and no values for Method 2.
(c) Hardness ratio (defined as the ratio of the 10--30~keV to 
3--10~keV absorbed fluxes), showing results from Methods 1 and 2.
(d) Blackbody temperature $kT$ for both Methods.  
(e) Blackbody radius $R_{bb}$ in km for both Methods, obtained
assuming a distance to the source of 8.0~kpc.
All panels:  hatched regions indicate epochs when Transient 1 (T1) and Transient 2 (T2) contaminated
our {\it NuSTAR} data; data for SGR J1745$-$2900 when T1 was on were unusable for flux or spectroscopy.
Vertical dotted red lines indicate epochs of X-ray bursts from the source direction.
\label{fig:spectral}
}
\end{figure}

The {\it NuSTAR} spectra are plotted in Figure~\ref{fig:all_spectra}.
We tested various spectral models for the magnetar, using \texttt{XSEPC} version 12.8.1.
We binned the spectra to have a minimum of 20 counts per spectral bin and used $\chi^2$ statistics in this method.
Physically, we do not expect the column density $N_H$
towards this source to vary on the relevant time scales, hence
we fit jointly for $N_H$ across all observations using the {\tt tbabs} function
in {\it XSPEC}.  The best
fit was obtained by the conventionally used empirical two-component model consisting of an absorbed blackbody
plus power law, with overall reduced $\chi^2_{\nu} = 1.03$ for 3683 degrees of freedom.
Our best-fit fluxes and blackbody temperatures and radii (calculated assuming a distance
of 8.0 kpc) are shown in Figure~\ref{fig:spectral}.
Fits with just one spectral component were significantly worse.
Although initially we allowed the blackbody and power-law
model parameters to vary, we also tried
fitting the models assuming a constant power-law index.  The resulting fits, with
constant and variable power-law index, are of similar
quality, and yield similar results for the blackbody component.  The best-fit value obtained for $N_H$, $(13.5 \pm 0.5) \times 10^{22}$~cm$^{-2}$,  
was consistent with the values reported by Mori et al. (2013) and Rea et al. (2013).  The best-fit value of the
assumed constant fit power-law index was $1.43 \pm 0.15$.  The best-fit $kT$ values and effective
radii are plotted in Figure~\ref{fig:spectral}.  The $\chi^2$ value for the overall fit was 6722
for 6425 degrees of freedom.  We note that our need for a power-law component, even a constant
value as a minimum, is consistent with the hint of high-energy excess reported by Rea et al. (2013) from
{\it Chandra} data.


Given that Transient 2 is $\sim$10 times brighter than the magnetar and that their PSFs overlap,
our method for handling the data affected by Transient 2 requires a special description.
Transient 2 was extracted and reduced in a similar fashion to the magnetar itself, 
but was fit with an absorbed blackbody plus disk model, based on similar fits
done to this source in the past \citep{yyt+09}.
The contamination of Transient counts within the magnetar extraction region was then estimated using the PSF
file from {\it NuSTAR}'s CALDB;  we concluded 0.035 of the transient's total flux
fell within our magnetar extraction region in our energy range, which amounted to roughly half
of the magnetar's flux in its aperture.  We then fit the spectra of both objects jointly 
but tying the Transient's parameters, and ignoring the magnetar's flux within the Transient's extraction region.
Allowing the contamination factor to vary gave similar results, and the best-fit contamination factor was close 
to the calculated one.  Note that we did not detect the source above 10 keV
in the last four {\it NuSTAR} observations so for those epochs 99\% confidence upper limits are presented
in the 10--30~keV energy range.  For all other observations, the detection significances in the 10--30 keV
band ranged from 10.4$\sigma$ (at the start of the observations) to 5.6$\sigma$ (at the last observation for which there
was a significant detection in this band).

In spectral analysis Method 2, 
we selected a background region
for each observation that included all of the chip on which the source
region falls, but excluded two bright features (Sgr A* and the Sgr A-E 
knot), as observed in a mosaic of all observations of the field in which
no transient was present, as well as stray light patterns from nearby
bright sources.  In individual observations, 
we found the selected background to always be subdominant compared to the
source spectrum in our energy range.
In this Method, we used \texttt{XSPEC} v.12.8.0m, binned the spectra
by a minimum of 3 counts per bin, and used the `lstat' fitting statistic.

We first fit a spectral model to the diffuse emission in the background
regions.  We jointly fit all the background regions, 
using two velocity-broadened APEC thermal plasma models
({\tt bapec} in \texttt{XSPEC}) plus a power-law component, with photoelectric absorption.
The velocity broadening of both plasma models was
linked, as were all diffuse emission parameters for all observations,
except the overall normalization, which was allowed to vary between modules A and B.
We fit the diffuse emission in observations where the magnetar is
absent, and found a reasonable model with featureless residuals throughout the fitted band.
This demonstrated to us that our background model is reasonable as a phenomenological
description sufficient for our purposes.  
For a more physically relevant consideration of the background
spectrum in this energy band, see \citet{ktb+13}.

We then fitted each magnetar observation individually using this independently
determined diffuse emission model with variable parameters as background.
For the magnetar's spectrum, we assumed a model consisting of a blackbody plus
power law, photoelectrically absorbed.
We used a Markov-Chain Monte Carlo method to explore the likelihood landscape
of a joint fit to all the data.
The absorption column of the magnetar's spectral model was linked for
all observations but the normalization was allowed to vary between modules
(though linked for each module between all observations), with all other
parameters left free to vary between observations (but linked for both
modules of the same observation).  Fitted power-law indexes were poorly
constrained.  When linking
the magnetar's power-law index across all observations, we found this to be slightly statistically disfavored but
the impact on the $kT$ values not significant.
Data contaminated by Transient 2 are not reported for Method 2 as for these we could not find stable results.

The results of our {\it NuSTAR} spectral analysis for the magnetar for both Methods are summarized
in Figure~\ref{fig:spectral}.  In the Figure, we show 
in the top two panels the flux evolution with time in two bands, 3--10 keV and 10--30 keV.
Note the difference in scale between the two in the Figure.
The results are qualitatively the same for Methods 1 and 2:  the soft band flux
decreases monotonically, quasi-linearly, by a factor of $\sim$2 over 80 days, while
the hard band flux shows greater variation -- notably an increase at the fourth
epoch 23 days post-outburst, and a greater decrease subsequently such that by 80 days post-outburst,
it is a factor of $\sim$4 lower than in our initial observation.
In panel (c) we plot a hardness ratio, defined as the ratio of the flux
in the 10--30~keV band to that in the 3--10~keV band. 
Methods 1 and 2 show
good qualitative and near-quantitative agreement: the source indeed hardens
significantly at the fourth epoch.  The third panel from the top shows the evolution
of blackbody temperature $kT$.
There is clear evidence for a decrease in $kT$ with time, at least until 60 days
post-outburst, in spite of the apparent increase in 10--30~keV flux at 23 days.
The bottom panel shows blackbody
radius evolution assuming a distance to the magnetar of 8.0 kpc.
Here we find our Methods disagree somewhat at the fourth
and fifth epochs; regardless, overall it is clear that the blackbody radius remained
nearly constant overall, to at least within 30--40\%.

The {\it Swift} spectral results described in \S\ref{sec:swift} are consistent with those of
{\it NuSTAR} in the 3--10 keV band, as shown in Figure~\ref{fig:spectral}, although we note a normalization
offset. This may be a result of cross-calibration uncertainties or due to imperfect background
subtraction, as we found the normalization of the {\it Swift} fluxes depended significantly
on the exact location of the selected background region.

\section{Discussion}

\subsection{Spin Evolution}

The coincidence of the second BAT burst with the intersection of the first
and second ephemerides, as shown graphically in Figure~\ref{fig:freq}, is striking.  
If not due to chance, the spin-down rate of the magnetar changed abruptly at the burst epoch
and there was no frequency discontinuity of any type, either a spin-up or spin-down glitch,
at that epoch.

Regardless of exactly how the change in spin-down rate occurred,
we have shown that its magnitude has
increased.  Comparing extrapolated values of $\dot{\nu}$ at the start of our observations
(MJD 56408) with that at the end (MJD 56519), we show that its magnitude has changed
by a factor of 2.60$\pm$0.07, that is, has nearly tripled in less than 4 months.  Moreover,
the magnitude is continuing to increase.

Large torque variations have been seen ubiquitously in magnetars
\citep[e.g.][]{wkg+02,kgw+03,gk04,ccr+07,crj+08,dksg12,akn+13,dk13}.  These torque
variations can be categorized in two main classes: (i) those following
glitches; and (ii) those unassociated with glitches.  
The former can be seen
immediately after a spin-up or spin-down event and are generally similar to the
recoveries seen post-glitch in many radio pulsars \citep[e.g.][]{ymw+10,elsm11,ymh+13}.
These recoveries are thought to be related to repinning of angular momentum vortices in
the superfluid component of the stellar crust after a major unpinning event
\citep{aaps84a,apc90,accp93}, although in the context of magnetars, scenarios
involving the magnetosphere have also been suggested \citep[e.g.][]{pbh12,pbh13,lyu13}.
\citet{tdw+00} discussed these possibilities in the context
of torque variations comparable to that seen in SGR J1745$-$2900 but observed in SGR 1900+14 
following its 1998 giant flare.  
However, since we find evidence for a glitch in SGR J1745$-$2900, 
glitch-specific models do not seem relevant here.  

\citet{tdw+00} proposed torque variations unassociated with glitches
could arise due to particle outflow. In this picture,
the energy output observed in the giant flare of SGR 1900+14 was insufficient
to explain the torque change; \citet{tdw+00} required a radiation-hydrodynamical outflow
of even higher energy output.  As no giant flare has
been observed from SGR J1745$-$2900, this model does not seem to apply here.

Magnetar torque variations have also been suggested to be purely magnetospheric in
origin \citep[e.g.][]{bel09}.
In general, a burst signals a sudden re-arrangement of a part
of the magnetosphere. If this part involves the open field lines,
the spin-down torque must change. 
Note that the open magnetic flux is a
tiny fraction $\sim10^{-4}$ of the total magnetic flux of the star, so the
spin-down torque is sensitive to the behavior of a tiny fraction of the
magnetosphere. 
In the context of the scenario proposed by \citet{bel09},
the persistent luminosity is produced by a much more
energetic closed j-bundle which may or may not contain the open flux. 
If a burst affects the open flux but does not affect the j-bundle much, the
torque can change while the X-ray luminosity does not, as is observed here, since
near the second burst epoch where the torque seems to have begun to change,
there is no feature in the source's flux evolution beyond the continued
decay following the source's initial appearance in 2013 April.
Indeed we find it interesting that at the second burst epoch, no glitch was seen
{\it and} no flux change was detected (apart from the brief burst itself and the continued decay following
the source's initial appearance).  It could
be that in magnetar outbursts, part of the observed enhanced flux results from 
the interior of the star and originates from heat released in an internal glitch, whereas another
component of the enhanced
flux originates from purely magnetospheric processes.
This could explain why, for example, in the 2002 outburst
of magnetar 1E 2259+586, there were two clearly different time scales associated
with the decay of the initial flux enhancement:  one very short, on a 1--2-day time scale,
and one lasting months \citep{wkt+04,zkd+08}.
Perhaps the short-term X-ray flux enhancement and decay is a result of transient glitch-related emission,
whereas the much longer decay is due to magnetospheric untwisting and/or 
crustal cooling.

Regardless of the origin of the torque variations, they clearly will dominate
over dynamical effects due to the orbit around Sgr A*.
\citet{rep+13} have argued that variations in the spin-down rate as
large as 10\% due to acceleration in the field of Sgr A* might one day be
observed, however, the observations reported here demonstrate that the
spin evolution of this magnetar, like those of many others, is inherently
unstable and unlikely to permit such a measurement.  The discovery of
a rotation-powered pulsar, particularly a millisecond radio pulsar,
in similar proximity to Sgr A*, would be far more useful for dynamical
studies of the black hole environment.

\subsection{Spectral Evolution}

We have shown that the 3--10 keV flux of SGR J1745$-$2900 has declined
since the source's initial appearance, albeit rather
slowly, only by a factor of $\sim$2 over the first 80 days since the discovery outburst.
This is similar to what was found by Rea et al. (2013) on the basis of three {\it Chandra}
observations.
The flux decay on a timescale of $\sim 10^7$~s was predicted by \citet{mgz+13} based on the 
observed emission area of the blackbody component, $A$, and the untwisting magnetosphere model of 
\citet{bel09}.  Specifically, \citet{mgz+13} estimated a luminosity evolution time scale of
$t_{ev} \simeq 10^7 \mu_{32} \Phi_{10}^{-1} A_{11.5}$~s, 
where $\mu_{32}$ is the magnetic moment in units of $10^{32}$~G~cm$^{3}$, 
$\Phi_{10}^{-1}$ is the electric voltage sustaining $e^{\pm}$ discharge in the magnetosphere 
in units of $10^{10}$~V,
and $A_{11.5}$ is in units of $10^{11.5}$~cm$^2$.  This predicted time scale is
roughly consistent with the flux decay we report.
The model also predicts that the hot spot should shrink, 
approximately as $A\propto L_{BB}^{1/2}$, where $L_{BB}$ is the blackbody luminosity. 
When $L_{BB}$ decreases by a factor of 2, area $A$ is expected to decrease by $\sim$40\% and 
the blackbody radius by $\sim$20\%. There may be a hint of such a radius 
decrease in the bottom panel of Figure 4, but given the contamination due to Transient 2,
our observations cannot confirm this.

The blackbody temperature also decreased monotonically by $\sim$20\% over
the first 60-days post-outburst, although as is clear in Figure~5, there is a possible hint
of an increase in the subsequent {\it NuSTAR} observation at 80 days.  Due to the presence
of Transient 2, we cannot verify unambiguously whether this trend continued, since our subsequent observations
may be contaminated and the {\it Swift} XRT observations yield insufficient statistics to detect
such a change.  {\it Chandra} observations, given that telescope's superior angular resolution
that should preclude the transient source contamination, may be able to address this question.  Meanwhile,
however, we note with interest the apparent hardening of the magnetar's flux in our fourth
observation 23 days post-outburst (see Fig. 5); this was unaccompanied by any significant frequency
or torque change, nor by any observed burst.  This is puzzling and
could indicate a burst that went unseen by all-sky monitors just prior, or perhaps it could be due
to a different source appearing within the {\it NuSTAR} PSF.
Regardless, we have not observed the common magnetar
flux/hardness correlation \citep[e.g.][]{wkt+04,gh07,zkd+08,tgd+08,re11,sk11} in this source; this may be due to the relatively
small range of fluxes yet observed, although we note no clear correlation was seen 
for SGR 1627$-$41 either, for a much larger flux range \citep{akt+12}.  Given the typical behavior of other magnetars, we expect the
source spectrum to gradually soften as the flux continues to decline, although presently, both flux
and temperature show evidence for leveling off.
We further note the relative stability of the inferred effective blackbody radius (Fig. 4);
any model to explain the flux decline will also have to account for a relatively stable
emitting area.
Some crustal cooling models predict an increase in emitting
area as the initially localized internal heat spreads around the neutron-star surface, while
relaxation following a magnetospheric twist as discussed above should involve a decrease in the emitting region, the
footpoint of the j-bundle \citep{bel09}.  Continued observations of the source spectrum
as the emission fades may help distinguish between these two possible processes in the star.

\section{Conclusions}

We have reported on X-ray observations made by {\it NuSTAR} and {\it Swift}
over $\sim$120 days after the initial outburst of the Galactic Center magnetar
SGR J1745$-$2900 in 2013 April.  We find that the magnetar's spin-down torque
has increased by a factor of nearly 3 compared with the spin-down rate initially
measured by \citet{mgz+13}, with no evidence for any accompanying spin-up or spin-down
glitch.  We also show that the pulsar's 3--10~keV flux has declined monotonically
by a factor of $\sim$2 over the first post-outburst 80 days, and that
the blackbody temperature has decreased by $\sim$20\% over the initial 60 days, similar
to what was reported by Rea et al. (2013),
although we find 
evidence for a possible levelling-off of both flux and temperature.  We observed a likely
increase in the source's 10--30 keV flux 17 days post-outburst, but observe
no accompanying timing or burst event.  We find no evidence for
the hardness/flux correlation commonly observed in magnetars, although this
seems likely due to the narrow range of fluxes we have yet sampled.  Further
monitoring may yet reveal spectral softening as the source flux declines.
We argue that the origin of the increase
in the spin-down rate is likely to be magnetospheric, and that such torque variations, 
ubiquitous in magnetars, are likely to dominate over any timing signatures of
motions related to the magnetar's proximity to Sgr A*.

This work was supported under NASA Contract No. NNG08FD60C, and made use of data from the {\it NuSTAR} 
mission, a project led by the California Institute of Technology, managed by the Jet Propulsion Laboratory, 
and funded by the National Aeronautics and Space Administration. We thank the {\it NuSTAR} Operations, 
Software and Calibration teams for support with the execution and analysis of these observations. This
research has made use of the {\it NuSTAR} Data Analysis Software (NuSTARDAS) jointly developed by the
ASI Science Data Center (ASDC, Italy) and the California Institute of Technology (USA).  
We acknowledge the use of public data from the Swift data archive.  This research has made use of the XRT 
Data Analysis Software (XRTDAS) developed under the responsibility of the ASI Science Data Center (ASDC), Italy.
We thank the {\it Swift} SOT team for their work in scheduling.
VMK receives support
from an NSERC Discovery Grant and Accelerator Supplement, from the Centre de Recherche en Astrophysique du Qu\'ebec, 
an R. Howard Webster Foundation Fellowship from the Canadian Institute for Advanced
Study, the Canada Research Chairs Program and the Lorne Trottier Chair
in Astrophysics and Cosmology.  RFA receives support from a Walter C. Sumner Memorial Fellowship.
AMB was supported by NASA grants NNX-10-AI72G and NNX-13-AI34G.
JAK was supported by supported by NASA contract NAS5-00136.
JKV's work was performed under the auspices of the U.S. Department of Energy by Lawrence Livermore National Laboratory under Contract DE-AC52-07NA27344. 


\newpage
\newcommand{\marka}{\tablenotemark{a}}
\newcommand{\markb}{\tablenotemark{b}}
\newcommand{\markc}{\tablenotemark{c}}
\begin{table}[t]
\begin{center}
\center{\caption{Timing Observations of SGR J1745$-$2900}}
\begin{tabular}{ccccc} \hline
Obs. ID  &  Date  & MJD\marka  & $T_{int}$ (ks) & Contamination\markb \\\hline
\multicolumn{4}{c}{{\it NuSTAR}} & \\\hline
30001002006 & 04/26/2013 & 56408 & 37.2 &   \\
80002013002 & 04/27/2013 & 56409 & 49.8  & \\
80002013004 & 05/04/2013 & 56416 & 38.6 &  \\
80002013006 & 05/11/2013 & 56423 & 32.7 &  \\
80002013008 & 05/18/2013 & 56430 & 39.0  & T1 \\
80002013010 & 05/27/2013 & 56439 & 37.4 & T1 \\
80002013012 & 06/14/2013 & 56457 & 26.7 & \\
80002013014/6 & 07/07/2013 & 56480 & 29.5\markc&  \\
80002013018 & 07/31/2013 & 56504  & 22.3 & T2   \\
80002013020 & 08/08/2013 & 56512 & 12.0  & T2\\
80002013022 & 08/09/2013 & 56513 & 11.2 & T2 \\
80002013024 & 08/13/2013 & 56517 & 11.7 & T2 \\\hline
\multicolumn{4}{c}{{\it Swift}} \\\hline
00032811001 & 05/03/2013 & 56415 & 15.5 \\
00032811002 & 05/11/2013 & 56423 & 9.2 \\
00032811003 & 05/16/2013 & 56428 & 9.5 \\
00032811005 & 05/19/2013 & 56431 & 13.5\\
0032811006 & 06/07/2013 & 56450 & 1.4\\
00032811008 & 07/15/2013 & 56488 & 12.6 \\
00032811009 & 07/16/2013 & 56489 & 1.0\\
00032811010 & 08/13/2013 & 56517 & 5.4 \\
00032811011 & 08/15/2013 & 56519 & 6.7 \\\hline
\end{tabular}
\tablenotetext{\rm a}{At the start of the observation.}
\tablenotetext{\rm b}{T1 is Transient 1; T2 is Transient 2; see \S 2.2.1 for details.}
\tablenotetext{\rm c}{The target fell in the stray light pattern of an unrelated source in Module B for
this observation only.  Hence for this observation B was omitted from the spectroscopic analysis.}
\label{ta:obs}
\end{center}
\end{table}

\newpage
\begin{table}[t]
\begin{center}
\caption{Phase-Coherent Timing Ephemerides for SGR J1745$-$2900}
\begin{tabular}{lc} \hline
Parameter  &  Value \\\hline
\multicolumn{2}{c}{First Ephemeris } \\\hline
MJD Range & 56408--56450 \\
Epoch & 56415.42 \\
Frequency $\nu$ (Hz) & 0.2657067288(20) \\
Frequency Derivative $\dot{\nu}$ (Hz) & $-4.32(9) \times 10^{-13}$   \\
Second Derivative $\ddot{\nu}$ (Hz/s) & $-8(1) \times 10^{-20}$   \\
Period $P\equiv 1/\nu$ (s) & 3.763547895(29) \\
Period Derivative, $\dot{P}$ & $6.12(12) \times 10^{-12}$ \\
Second Derivative, $\ddot{P}$ (s$^{-1}$) & $1.15(15) \times 10^{-18}$    \\
RMS Residual (ms) & 38 \\
$\chi^2$/dof/p\marka & 113/86/0.03 \\\hline
\multicolumn{2}{c}{Second Ephemeris } \\\hline
MJD Range & 56457--56519 \\
Epoch &  56513.00 \\
Frequency $\nu$ (Hz) & 0.265700350(9) \\
Frequency Derivative $\dot{\nu}$ (Hz) & $-9.77(10) \times 10^{-13}$   \\
Second Derivative $\ddot{\nu}$ (Hz/s) & $-2.7(4) \times 10^{-20}$   \\
Period $P\equiv 1/\nu$ (s) & 3.76363824(13) \\
Period Derivative, $\dot{P}$ & $1.385(15) \times 10^{-11}$ \\
Second Derivative, $\ddot{P}$ (s$^{-1}$) & $3.9(6) \times 10^{-19}$    \\
RMS Residual (ms) & 51 \\
$\chi^2$/dof/p\marka & 52/41/0.12 \\\hline
\hline
\end{tabular}
\tablenotetext{\rm a}{$\chi^2$, degrees of freedom, probability of chance occurrence.}
\label{ta:timing}
\end{center}
\end{table}

\clearpage

\clearpage

\clearpage

\clearpage

\clearpage

\end{document}